\documentclass[conference]{IEEEtran}
\usepackage{adjustbox}
\usepackage{todonotes}
\usepackage{xcolor}
\definecolor{darkgreen}{rgb}{0.0, 0.5, 0.0}
\usepackage{multirow}
\usepackage{amsmath,amssymb,amsfonts}
\usepackage{caption}
\usepackage{textcomp}
\usepackage{xcolor}
\usepackage{svg}
\usepackage{afterpage}
\usepackage{booktabs}
\usepackage{float}
\usepackage{subfigure}
\usepackage{multirow}
\usepackage{todonotes}
\usepackage{soul}
\usepackage[citecolor=blue,colorlinks=true,linkcolor=blue]{hyperref}%
\usepackage{enumitem} 
\ifCLASSINFOpdf
\else
\fi
\begin{document}
%
\title{SecCAN: An Extended CAN Controller with Embedded Intrusion Detection}


\author{\IEEEauthorblockN{Shashwat Khandelwal \& Shanker Shreejith}
\IEEEauthorblockA{ Reconfigurable Computing Systems Lab, Electronic \& Electrical Engineering \\
Trinity College Dublin, Ireland\\
Email: \{khandels,shankers\}@tcd.ie}}


%


\maketitle

\begin{abstract}\label{sec:abstract}
Recent research has highlighted the vulnerability of in-vehicle network protocols such as controller area networks (CAN) and proposed machine learning-based intrusion detection systems (IDSs) as an effective mitigation technique.
However, their efficient integration into vehicular architecture is non-trivial, with existing methods relying on electronic control units (ECUs)-coupled IDS accelerators or dedicated ECUs as IDS accelerators. 
Here, initiating IDS requires complete reception of a CAN message from the controller, incurring data movement and software overheads.
In this paper, we present SecCAN, a novel CAN controller architecture that embeds IDS capability within the datapath of the controller.
This integration allows IDS to tap messages directly from within the CAN controller as they are received from the bus, removing overheads incurred by existing ML-based IDSs.
A custom-quantised machine-learning accelerator is developed as the IDS engine and embedded into SecCAN's receive data path, with optimisations to overlap the IDS inference with the protocol's reception window.
We implement SecCAN on AMD XCZU7EV FPGA to quantify its performance and benefits in hardware, using multiple attack datasets. 
We show that SecCAN can completely hide the IDS latency within the CAN reception window for all CAN packet sizes and detect multiple attacks with state-of-the-art accuracy with zero software overheads on the ECU and low energy overhead (73.7$\mu$J per message) for IDS inference.
Also, SecCAN incurs limited resource overhead compared to a standard CAN controller ($<$\,30\%\,LUT, $<$\,1\%\,FF), making it ideally suited for automotive deployment.
\end{abstract}
\section{Introduction and Background}\label{sec:introduction}
Most high-end vehicles today integrate over 50 electronic computing units (ECUs) interconnected through different network standards for incorporating safety-critical, comfort and automation capabilities in a cost and energy-efficient manner. 
CAN (and its variants) continue to be the most widely used network protocol in automotive electric/electronic systems owing to their low cost, flexibility, and robustness~\cite{hpl2002introduction}. 
As a broadcast-based shared-bus protocol with minimal overhead, CAN has no built-in scheme for securing message exchanges over the network or providing sender/receiver authentication. 
Thus, any malicious node accessing the physical network can easily observe, decode and tamper CAN messages~\cite{rajapaksha2023ai}.

\begin{figure}[t!]
    \centering
    \includegraphics[scale = 0.48, trim={0.7cm 0 0.75cm 0},clip]{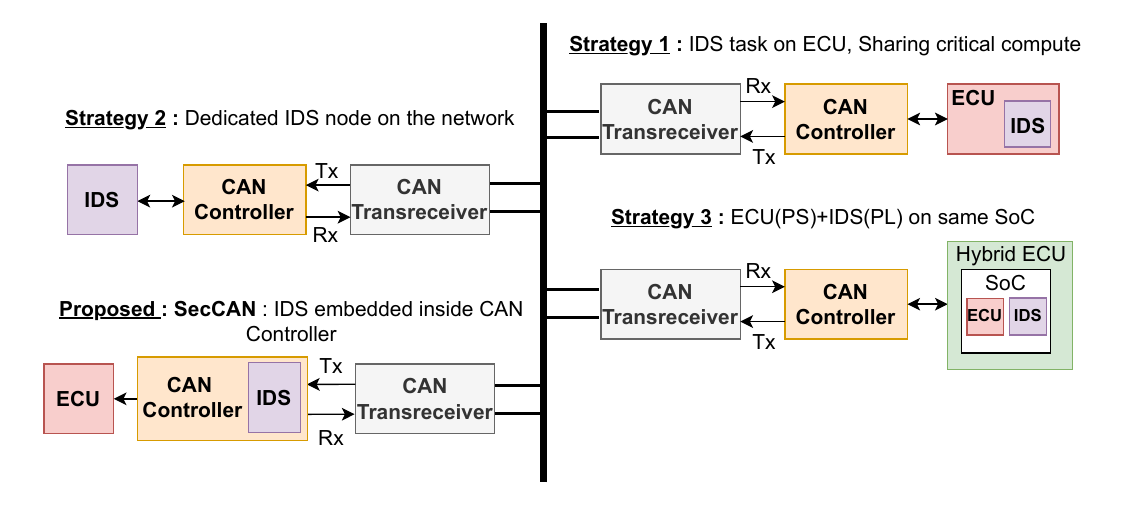}
    \caption{The figure illustrates conventional integration strategies for CAN IDSs reported in the literature, and the proposed case for embedding IDS within the controller.}
    \label{fig:ids-integration} \vspace{-6mm}
\end{figure}

To mitigate these vulnerabilities, researchers have explored methods to detect and restrict malicious actors using intrusion detection and intrusion prevention systems (IDSs and IPSs respectively)~\cite{bhatia2021evading}.
Compared to rule-based and entropy-based IDS systems, the generalisable and scalable nature of ML-based IDS has made them ideally suited for embedded intrusion detection in vehicular ECUs~\cite{narayanan2015using,alshammari2018classification,yang2019tree,song2020vehicle,tariq2020cantransfer}.
However, the computational complexity of ML-based IDS is prohibitive to software deployment on the ECU as the IDS task can consume valuable time and resources required for the safety-critical and real-time functions on the ECU. 
Hence, most practical IDS integration schemes in the literature rely on complete ECUs dedicated to IDS (GPUs, FPGAs, or microcontrollers) or through a dedicated accelerator attached to the ECU.
Figure~\ref{fig:ids-integration} captures the different integration strategies proposed in the literature. 
Case 1 in figure~\ref{fig:ids-integration} shows the IDS integrated as a software task on an existing ECU, while case 2 shows a dedicated IDS accelerator where the IDS could be deployed as the lone software task on an ECU, GPU (edge device or a standard GPU) or a microcontroller platform like Raspberry Pi~\cite{9654211}. 
Case 3 shows a coupled accelerator, where the ECU offloads IDS to the accelerator on the same SoC once the packet is received from the CAN controller.
In all the above cases, the CAN message has to be completely received by the controller and subsequently read by the coupled ECU/processing element before the ML model can perform IDS checks. 

In this letter, we propose SecCAN, an extended CAN controller that integrates an IDS accelerator into its receive side datapath, as shown in figure~\ref{fig:ids-integration} case 4. 
This critical design choice allows the IDS accelerator to directly extract CAN bus data (ID and payload) from within the receive path of the CAN controller for overlapping IDS execution with the current packet's reception. 
Complementing this with a compact 4-bit quantised multi-layer perception (QMLP) IDS model deployed as an unrolled dataflow accelerator, SecCAN completely hides the latency of IDS within the reception window of the current CAN frame.
This is a departure from existing ML-based IDS solutions in the research literature, where IDS execution is triggered after the message is transferred to the ECU from the CAN controller.
We evaluate SecCAN controller on a Zynq Ultrascale+ platform with the ARM cores on the Zynq device acting as the coupled ECU.
The IDS performance of the SecCAN controller is evaluated by replaying messages from multiple datasets on the test platform.
The key contributions of the letter are as follows:
\begin{itemize}[leftmargin=*]
    \item We introduce a novel CAN controller architecture (SecCAN) that seamlessly integrates IDS capabilities into the datapath of the controller without affecting standard message flow, thus being fully transparent to the ECU.
    \item A lightweight 4-bit QMLP model was developed and dataflow optimised to generate the low-latency quantised IDS (Q-IDS) accelerator integrated into SecCAN.
    The model was trained and tested on two attack datasets~\cite{song2020vehicle,han2018anomaly} and achieved state-of-the-art binary classification accuracy of 99.993\% and 99.966\% (average) across multiple attacks. 
    \item We implement the SecCAN controller on an AMD Zynq XCZU7EV FPGA to determine the latency, area and energy benefits/overheads. The tests show that SecCAN enables line-rate IDS capability with zero latency and zero software overheads incurred by the ECU even at the highest CAN data rates. Compared to dedicated IDS accelerators such as on a Jetson Xavier GPU, SecCAN is significantly more energy-efficient (34$\times$) with the IDS implementation in SecCAN incurring much lower latency (10.9$\times$).
\end{itemize} 
The datapath changes in SecCAN allows IDS to be fully offloaded from the ECU, with clever overlapping of IDS with the message reception from the CAN bus. 
This makes IDS fully transparent to the ECU, effectively incurring zero latency for the IDS and with zero change to existing software tasks, a key difference to competing ML-based IDS methods~\cite{song2020vehicle, yang2021mth, ma2022gru, agrawal2022novelads}. 
To the best of our knowledge, this work presents the first integration of an IDS operating in parallel with the CAN controller’s datapath.
%
We further openly release the implementation of SecCAN to foster further research into smart network interfaces at \href{https://github.com/RCSL-TCD/SecCAN}{https://github.com/RCSL-TCD/SecCAN}.



\section{SecCAN Architecture}\label{sec:mlfpga}
\subsection{Extending Datapath for IDS}
\begin{figure}[t!]
    \centering
    \hspace{-0.5cm}
    \includegraphics[scale = 0.56,trim={0.0cm 0 0.6cm 0},clip]{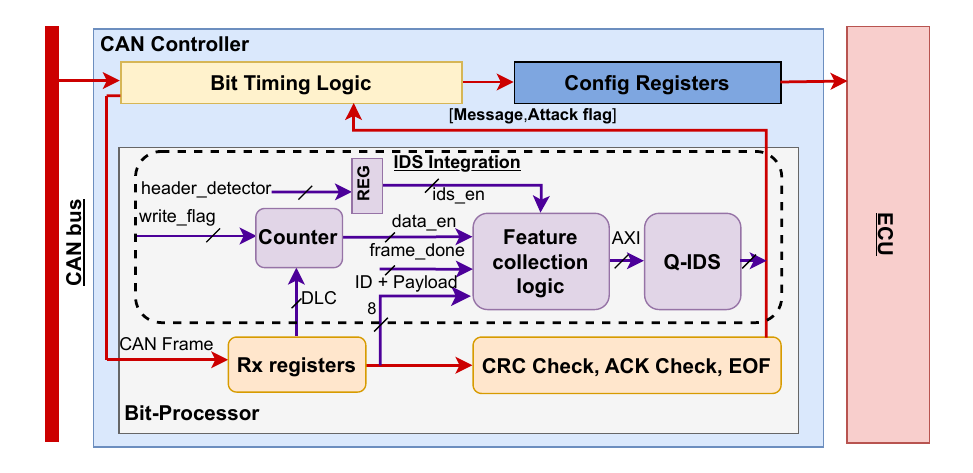}
    \caption{The figure illustrates the integration of the IDS within the CAN controller. The red arrows show the standard datapath and the purple arrows indicate the augmented path for IDS.}
    \label{fig:can_bsp} \vspace{-6mm}
\end{figure}

The CAN controller implements the CAN protocol at the logic level using functional blocks that perform bit-level processing, data packaging/framing, error detection, and optional message filtering functions. 
For our work, we utilise an open-source register transfer level (RTL) implementation of the CAN controller described in Verilog as our baseline controller~\cite{opencorecan}.
The CAN protocol is implemented in the baseline CAN controller through the following modules: \textit{config\_registers}, \textit{bit\_timing\_logic} and \textit{bit\_processing}. 

The \textit{config\_registers} module implements the AXI interface to the ECU and incorporates a set of registers and frame buffers to capture protocol configuration, status signals and transmit/received messages. 
The memory-mapped registers within the \textit{config\_registers} module define command/control functions for other modules as well as error-tracking and status signals. 
At startup, the ECU configures the protocol registers to set operating parameters such as bit-level timing (sampling windows, clock prescaler configuration), interrupt configuration, and message filtering for the receive path, among others. 
Once active, a software task in the ECU writes up to 8 bytes of data to be sent as a CAN frame into the transmit buffer, and the controller attempts to transmit the payload based on the configured arbitration priority. 
The status of transmission is subsequently updated on the status register. 
When a CAN frame is received from the bus, the message filter configuration determines if it needs to be passed to the ECU or dropped by the controller. 
Any frame to be passed upstream is written to the receive buffer and an interrupt signal is generated (if not masked), causing the ECU to read the received frame. 


The \textit{bit\_timing\_logic} module implements the bit-level timing and synchronization functions of the protocol and interfaces with the physical CAN bus.
The \textit{bit\_processing} module processes the encoding/decoding of bits to/from the CAN bus and implements the transmit, receive and error handling functions.
We extend the datapath within the \textit{bit\_processing} module to embed our 4-bit quantised MLP as the IDS (Q-IDS) within the CAN IP, as shown in figure~\ref{fig:can_bsp}.
The extension brings together a set of control signals and the decoded byte from the controller's modules to the feature collection logic that generates the input features for our IDS. 
The Q-IDS IP is an AXI stream accelerator of our quantised MLP generated using AMD's FINN toolchain, starting from our high-level Python model. We apply selective unrolling and dataflow optimisations to optimise throughput and latency in the generated IP.

\begin{figure*}[th!]
    \centering
    \includegraphics[scale = 0.6,trim={0.8cm 0 0.0cm 0},clip]{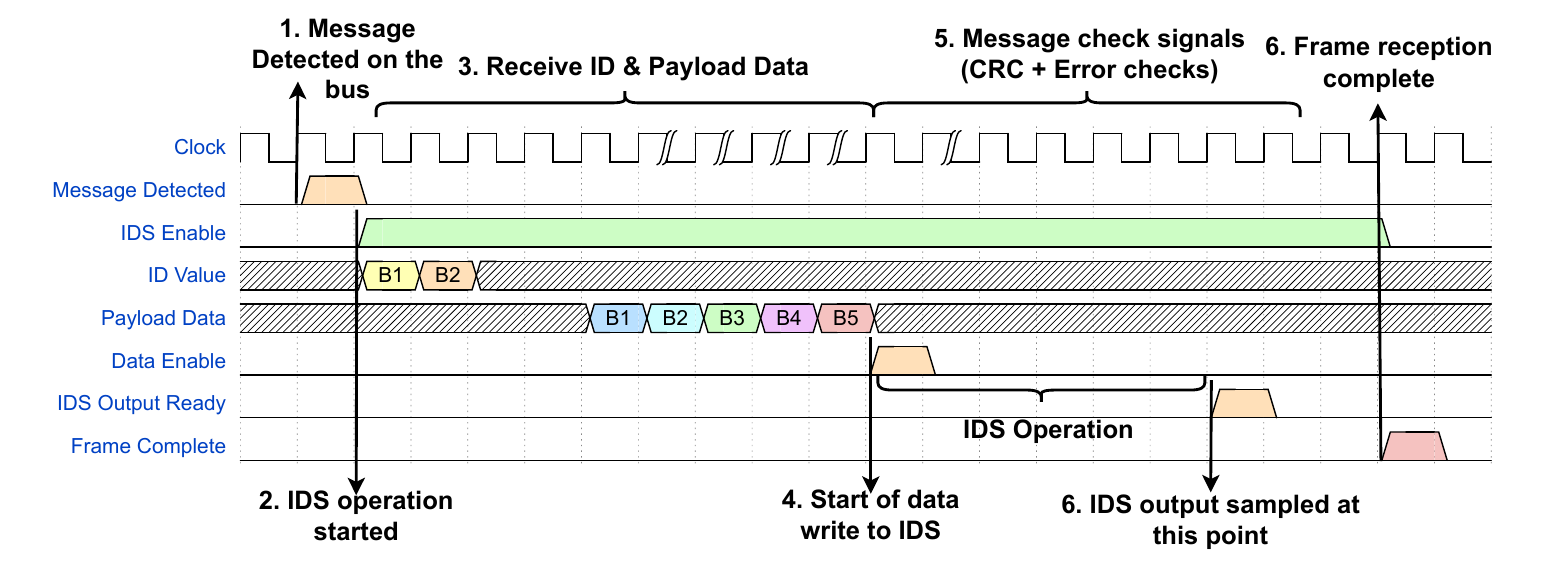}
    \caption{The waveform shows the signalling within the SecCAN controller for a 5-byte CAN message from the bus. The IDS operation is overlapped with the protocol checks on the bus and is completed before the frame is ready to be read by the ECU.}
    \label{fig:waveform} \vspace{-6mm}
\end{figure*}

The bit processing operation begins when a header is detected on the bus and is indicated by the \textit{header\_detector} signal, marking a possible start of a new message on the CAN bus. 
The IDS control logic asserts \textit{ids\_en} to enable the feature collection logic and the counter starts tracking the number of data bytes written to the receive data register(s) (and also to our feature collector FIFO through the datapath extension) by monitoring the \textit{write\_flag} signal. 
When the counter value matches the data length code (DLC) of the current CAN frame, the frame header and message are available in the feature collection logic's FIFO, and the \textit{data\_en} signal is asserted to begin the IDS operation. 
The new input feature vector comprising (CAN ID $+$ Payload) of current and previous messages is subsequently transferred to the IDS IP.  
The feature collection logic zero pads smaller-sized payloads to 8 bytes to generate a uniform feature size for the IDS. 
This allows us to replay the bus messages as-is during our evaluation without pre-processing the dataset, faithfully replicating an in-vehicle scenario. 
Once the feature vector is transferred, the IDS processes the frame for potential attacks and asserts the \textit{ids\_output\_ready} signal on completion.   
This operation overlaps with the CAN protocol checks, where the bit-processor validates the CRC and error flags of the received frame, and waits for the end-of-frame signal before transferring the valid CAN message to the receive buffer (in \textit{config\_registers} module). 
The IDS output value is wired out of the \textit{bit\_processing} module to the \textit{top} module, where a custom multiplexing logic appends the IDS output to the received CAN message as it is transferred to the receive buffer and subsequently read by the ECU. 
Figure~\ref{fig:waveform} illustrates the SecCAN operation using a waveform, highlighting how the IDS operations overlap with the CAN signalling checks (CRC, error flagging bit times). 
Our optimised IDS implementation completes the inference within this window, thus hiding this latency from the ECU.

\subsection{Model \& Dataset for training and testing}\label{subsec:MLmodel}
To arrive at our final QNN model's architecture, we explored different configurations with varying complexity (number of layers and number of neurons in each layer) to find a model that offers high inference accuracy at minimal complexity. 
We use CAN ID $+$ Payload information from each CAN frame as the input feature for the model to perform binary classification. 
We arrived at a 4-bit (weights/activations) quantised multilayer perceptron model (MLP) as the chosen configuration for the IDS, which provided the best validation accuracy during the training process.
The QNN model is trained using \emph{brevitas}, which is an open-source quantisation-aware training (QAT) library for training neural networks~\cite{brevitas}.
The model was trained for 200 epochs with the \textit{adam} optimizer and \textit{lr} set to 0.0001.
The model consists of 3 \emph{QuantLinear} layers (input, one hidden \& output ; \emph{Linear} layer equivalent in pytorch)  with \emph{\{64, 32, 1\}} neurons at each layer, followed by \emph{QuantRelu} activations. 
Batch normalisation and dropout layers were used to prevent overfitting during the training phase. 
The sigmoid function at the output denotes the probability of the current message being benign or malicious. 

We used the open Car Hacking \& survival analysis dataset for training and testing the model~\cite{song2020vehicle,han2018anomaly}.
The datasets contain CAN bus data acquired via the OBD port in an actual vehicle, with attack messages injected in real-time, compiled into a labelled set of normal and attack messages.
We split each dataset into 75:15:10 proportions for training, validation, and testing. 

\section{Evaluation \& Results}\label{sec:experiments}
We generate the hardware configuration of the model using AMD Vivado 2022.2, with the XCZU7EV device on the ZCU104 development board as the FPGA platform.
The CAN controller is set to operate at maximum bitrate of 1 Mbps, with a CAN clock frequency of 16 MHz. 
We evaluate the 4b-QMLP IDS model's accuracy in detecting active injection attacks and compare it to state-of-the-art approaches in the literature.
The 4b-QMLP model uses the same 16 MHz to avoid  clock domain crossing; however, it can be synthesized for a much higher clock frequency (100\,MHz) if required at the cost of higher resource and energy consumption (see sec.~\ref{subsec:overheads}).
Additionally, we measure the inference latency in hardware and analyse the trade-offs in resource and energy utilisation that the controller incurs from this integration.
For measurements, we replay the attack messages either directly from the ARM processor core (on the Zynq platform for testing accuracy) or from a BRAM that replays CAN traffic data (for quantifying latency and power).

\subsection{Accuracy}
We quantify the accuracy of the 4b-QMLP model in terms of precision, recall \& F1 scores and compare it against state-of-the-art works in the literature, and across multiple datasets.  
Our test split from the Car Hacking dataset incorporates 75,000 messages each from the DoS \& fuzzing attack datasets.
The model achieves a binary classification accuracy of 99.993\% (\emph{11 misclassifications for 150,000 test messages}),
Our test split from the survival analysis dataset combines flooding, fuzzing, and malfunction attack vectors. 
On this test set, our model achieves a classification accuracy of 99.966\% (\emph{25 misclassifications for 75,000 test messages}).
Table~\ref{table:ch-accuracy} presents a comparison of our model with others proposed in the literature for the two attack datasets, highlighting that our approach matches or surpasses the performance of competing methods.

\begin{table}[b!]
\centering
\caption{Inference accuracy (\%) of SecCAN compared to competing IDS schemes on both datasets.}
\scalebox{0.85}{
\begin{tabular}{@{}llllll@{}}
\toprule
\textbf{Attack}  & \textbf{Model} & \textbf{Precision} & \textbf{Recall} & \textbf{F1}  & \textbf{\% FNR} \\
\midrule
\multirow{6}{1.0cm}{DoS (Car Hacking)} 
& DCNN~\cite{song2020vehicle}  & 100 & 99.89  & 99.95  & 0.13 \\
& NovelADS~\cite{agrawal2022novelads} & 99.97 & 99.91 & 99.94  & -  \\
& TCAN-IDS~\cite{cheng2022tcan} & 100  & 99.97  & 99.98  & -    \\
& GRU~\cite{ma2022gru} & 99.93 &   99.91  &  99.92 &  -   \\ 
& iForest~\cite{9654211}   & 95.07     & 99.93   & 97.44  &  - \\
& \textbf{4b-QMLP in SecCAN}  &      99.99  & 99.98  & 99.98  & 0.02     \\
\midrule
\multirow{6}{1.0cm}{Fuzzing  (Car Hacking)}
& DCNN~\cite{song2020vehicle} & 99.95 & 99.65 & 99.80  & 0.5  \\
& NovelADS~\cite{agrawal2022novelads} & 99.99 & 100 & 100  & -    \\
& TCAN-IDS~\cite{cheng2022tcan} & 99.96 & 99.89 & 99.22  & -    \\
& GRU~\cite{ma2022gru} & 99.32 & 99.13 & 99.22  &    - \\
& iForest~\cite{9654211} & 95.07 & 99.93 & 97.44  &    - \\
& \textbf{4b-QMLP in SecCAN} &99.99  & 99.97  &  99.98  &   0.03   \\
\midrule\midrule
\multirow{4}{1.0cm}{Flooding (Survival Analysis)} 
& XGBoost~\cite{9797224}               &   100 & 90 & 94.74 & -  \\
& G-IDCS~\cite{park2023g}         & 99.72 & 99.72 & 99.72 & -\\
& LSTM~\cite{9216166}                 & - & 100 & 100 & 0\\
& \textbf{4b-QMLP in SecCAN}  &    100& 100  & 100 & 0  \\
\midrule
\multirow{4}{1.0cm}{Fuzzing (Survival Analysis)}
& XGBoost~\cite{9797224}                & 99.98 & 99.08 & 99.53 & - \\
& G-IDCS~\cite{park2023g}       & 100 & 100 & 100 & 0\\
& LSTM~\cite{9216166}                  & - & 99.95 & 99.96 & 0.05\\
& \textbf{4b-QMLP in SecCAN} &99.98      &     99.50    & 99.74 & 0.5\\
\midrule
\multirow{4}{1.2cm}{Malfunction (Survival Analysis)}
& XGBoost~\cite{9797224}                & 99.92 & 100 & 99.96 & - \\
& G-IDCS~\cite{park2023g}        & 100 & 99.64 & 99.82 & - \\
& LSTM~\cite{9216166}                & - & 100 & 100 & 0 \\
& \textbf{4b-QMLP in SecCAN} &100     &  100         & 100  & 0 \\
\bottomrule
\end{tabular}}
\label{table:ch-accuracy}
\end{table}

\subsection{Detection Latency}
For real-time message tagging, the IDS must complete the analysis of each message before the reception window is completed (end of error flags in CAN protocol). 
For a maximal-length CAN message, the reception window \textit{${T_{max}}$} can be determined as  \( T_{max} = T_{frame\_done} - T_{header\_detected} \), where \textit{$T_{frame\_done}$} \& \textit{$T_{header\_detected}$} marks the time at completion of frame reception and valid header reception respectively. 
Since our IDS uses header and payload as input features, the analysis can only start once all data bytes are decoded from the bus (indicated by \text{${data\_en}$} signal). 
For CAN bus operating at 1\,Mbps, this time window can be determined as 37.376\,\textmu s; hence, for real-time detection, the upper limit on IDS latency  (\textit{${T_{IDS}}$}) must satisfy \(T_{IDS} = T_{frame\_done} - T_{data\_en}\,<\,37.376\,\mu s\).
This relation holds for any data length at 1\,Mbps CAN bus configuration. 
From our evaluation (both from simulation and hardware measurements), we observe a detection latency of 36.5\,\textmu s at 16\,MHz clock, allowing the SecCAN to tag each message with an attack or benign flag before the reception window of current message on the network ends.
Further, we observe a 10.9$\times$ reduction in latency for our unrolled 4b-QMLP compared to an 8-bit implementation on a Jetson Xavier, even with reception and message transfer times excluded.

\subsection{Resource \& Energy Consumption}\label{subsec:overheads}
We report the resource consumption and energy numbers of the SecCAN controller, compared to the standard CAN controller in table~\ref{table:resourceutilization} \& ~\ref{tab:seccan_comparison} respectively.
We can observe that the IDS model contributes $\approx$ 29.5\% additional LUTs, $\approx$ 0.45\% additional FFs and $\approx$ 60.5\% LUTRAMs over the standard CAN controller. 
The higher LUT/LUTRAM usage results from a fully unrolled implementation of the model, which allows the model to achieve line-rate analysis at the same clock rate as the controller. 
We further measure the per-message energy consumption of the IDS by monitoring the power rails on the ZCU104 board during IDS execution.
Table~\ref{tab:seccan_comparison} compares SecCAN's energy consumption per inference against other works in the literature which have reported energy consumption.
Our model consumes only 73.7$\mu$J per message on average, a 3.1$\times$ improvement over our 2-bit quantised coupled accelerator IDS~\cite{10296211}, achieved primarily through lower operating speed and the Q-IDS IP optimisations.
Compared to software IDS, SecCAN achieves a 34.1$\times$ reduction in energy consumption over an 8-bit variant of our model on a Jetson Xavier. 
Similarly, SecCAN consumes 24.1$\times$ and 28.4$\times$ lower energy per inference than a GRU-based IDS~\cite{ma2022gru} and a convolutional autoencoder-model~\cite{khandelwal2023real} on Jetson Xavier and Zynq Ultracale+ platforms respectively. 
While other competing schemes such as MTH-IDS~\cite{yang2021mth} and iForest~\cite{9654211} do not explicitly report their energy consumption, they are re-implemented on Raspberry Pi-3 and Pi-4 devices for comparison (as in the original article). 
In comparison to these, SecCAN's energy consumption was found to be 17.6$\times$ and 5.3$\times$ lower per inference.
It should be noted that energy measurement for SecCAN is for the extended CAN controller (SecCAN) as opposed to the IDS-only energy consumption reported and measured for competing schemes.

\begin{table}[t]
\caption{Resource utilisation on the FPGA for the standard CAN controller (CAN-NC) \& our SecCAN controller.}
\begin{center}
\scalebox{0.9}{%
\begin{tabular}{@{}lcccc@{}}
\toprule
\textbf{} & \textbf{LUTs} (\%) & \textbf{FFs} (\%) & \textbf{BRAMs} (\%)  & \textbf{LUTRAMs} (\%) \\
\midrule
\midrule
CAN-NC & 887 (0.38\%) & 625 (0.14\%)  & 0 (0) & 18 (0.02\%)  \\
4b-QMLP & 67902 (29.47\%) & 2007 (0.44\%)  & 0.5 (0.16\%) & 61482 (60.42\%)\\
SecCAN & 68888 (29.90\%) & 2737 (0.59\%) & 0.5 (0.16\%) & 61500 (60.44\%) \\
\bottomrule \vspace{-4mm}
\end{tabular}}
\label{table:resourceutilization}
\end{center} 
\end{table} 

\begin{table}[t!]
\centering
\caption{Comparison of SecCAN’s energy consumption per inference with other IDS approaches proposed in the literature.}
\scalebox{0.95}{%
\begin{tabular}{@{}llc@{}}
\toprule
\textbf{Model} & \textbf{Platform} & \textbf{Energy consumption} \\ \midrule
GRU~\cite{ma2022gru} & Jetson Xavier NX & 1.77 $m$J \\
QCAE~\cite{khandelwal2023real} & Zynq Ultrascale+ & 2.09 $m$J\\
MTH-IDS~\cite{yang2021mth} & Raspberry Pi 3  & 1.29 $m$J \\
iForest~\cite{9654211} & Raspberry Pi 4 & 390.6 $\mu$J\\
\textbf{SecCAN} & Zynq Ultrascale+ & 73.7 $\mu$J\\
\bottomrule
\end{tabular}
\label{tab:seccan_comparison}}
\vspace{-2mm}
\end{table}
\section{Conclusion}\label{sec:conclusion}
In this letter, we explore a smart CAN controller architecture that integrates a light-weight machine learning model as an IDS accelerator within the CAN controller (SecCAN controller) to detect the onset of intrusions from CAN traffic flow. 
By integrating the IDS within the controller, the proposed SecCAN controller can classify benign/attack messages before the reception window is completed—an approach that has not been explored before to the best of our knowledge. 
This flow allows safety mechanisms and intrusion prevention systems to be triggered as soon as the ECU receives a message, unlike conventional integration approaches where the IDS processing can begin only after the ECU has fully received the message from the CAN controller.
The embedded IDS consumes only 73.7$\mu$J per classification of each message while achieving 99.993\% \& 99.966\%  detection accuracy across multiple attack vectors and datasets, with a low resource overhead of $<$ 30\% LUT \& $<$ 1\% FF on an AMD Zynq XCZU7EV. 
We believe that the SecCAN architecture is scalable and can pave the way towards smarter IDS approaches in existing and future automotive networks.
\IEEEpeerreviewmaketitle






%



\bibliography{references} 
\bibliographystyle{ieeetr}

\end{document}